\documentclass[pre,onecolunm]{revtex4}

\usepackage{tikz}
\usepackage{amsmath}

\begin{document}

	\title[Local vs. long-range infection in 1d epidemics]{Local vs. long-range infection in unidimensional epidemics} 
\author{Priscila R. Silveira}
\author{Marcelo M. de Oliveira}\email{mmdeoliveira@ufsj.edu.br}
\author{Sidiney G. Alves}\email{sidiney@ufsj.edu.br}
\affiliation{Departamento de F\'{\i}sica e Matem\'atica,
	CAP, Universidade Federal de S\~ao Jo\~ao del Rei,
	36420-000 Ouro Branco, Minas Gerais - Brazil}

	\begin{abstract}
		
		We study the effects of local and distance interactions in the unidimensional contact process (CP). 
		In the model, each site of a lattice is occupied by an individual, which can be healthy or infected.
		As in the standard CP, each infected individual spreads the disease to one of its first-neighbors
		with rate $\lambda$, and with unitary rate, it becomes healthy. However, in our model, an infected  
		individual can transmit the disease to an individual at a distance $\ell$ apart. This step mimics a vector-mediated 
		transmission. We observe the host-host interactions do not alter the critical exponents significantly in comparison to a process with only L\'evy-type interactions. Our results confirm, numerically, early field-theoretic predictions. 
	\end{abstract}
	
	{\tiny}

	\maketitle

	\section{Introduction}
	
	The contact process (CP) \cite{harris-CP}
	is a stochastic epidemics model with spatial structure \cite{durrett}.
	In the CP, each individual inhabits a site on a $d-$dimensional lattice and can be in one of
	two states: healthy or infected. Each infected individual transmits the disease to one of its
	nearest-neighbors with rate $\lambda$, or become healthy with a unitary rate.
	When the transmission rate $\lambda$ is varied, the system undergoes a phase
	transition between disease-free and endemic phases.
	
	Apart from its interest as an elementary spatial model of epidemic spreading, 
	the critical behavior of the CP (and its variations) is interesting in the study of 
	nonequilibrium universality classes. The disease-free state is an absorbing state, 
	a frozen state with no fluctuations \cite{marro,henkel,odor07,hinrichsen,odor04}.
	Nonequilibrium phase transitions into absorbing states have been a topic of much interest in recent decades. 
	In addition to their connection with
	epidemics, they appear in a wide variety of problems, such as heterogeneous catalysis \cite{zgb},
	interface growth \cite{tang}, population models and ecology \cite{scp}. 
	Recent experimental realizations in the liquid crystal
	electroconvection \cite{take07}, driven suspensions \cite{pine} and superconducting vortices \cite{okuma}
	have heightened interest in absorbing transitions. 
	
	It is expected  that absorbing state phase transitions in
	models with a positive unidimensional order parameter,
	short-range interactions, and without additional symmetries
	or quenched disorder belong generically to the universality
	class of directed percolation (DP)\cite{janssen, grassberger}.
	Including long-range interactions in spreading processes can provide more realistic 
	models, instead of short-range models as the original CP, for example, to model spreading of vector-borne diseases \cite{dengue}. One of the first approaches was the 
	model proposed by Grassberger \cite{grass86}, based on the original idea presented in \cite{mollison}. In his model,
	the infection probability obeys a L\'evy flight decaying as a power-law relation $1/r^{\alpha+d}$
	with the distance $r$, where
	$d$ is the spatial dimension of the system and $\alpha$ is a control
	parameter. Simulational \cite{hinri99,hinri07} and field-theoretical renormalization group analysis \cite{janssen99} revealed that such {\em anomalous} directed
	percolation presents critical exponents varying continuously with the
	parameter $\alpha$. 
	
	A generalized version of the unidimensional CP where inactive (healthy) sites can be activated (infected) over long distance was
	introduced by Ginelli et al. \cite{ginelli05,ginelli06}, inspired by
	pinning-depinning transitions in nonequilibrium wetting
	phenomena. In such model, an active site infects an inactive with rate $q/\ell^\alpha$, where $q$ is the coordination number and $\ell$ is the distance between both sites. 
	Depending on the value of the control parameter $\alpha$, the contact process with long-range interactions exhibits a rich
	phase diagram, with distinct universality classes and discontinuous
	phase transitions \cite{ginelli05,ginelli06,hinri07,fiore07}. The robustness of these discontinuous phase transitions was studied in \cite{fiore13}. More recent studies focused on the effects of quenched disorder \cite{igloi} and diffusion \cite{pedro}.  Inspired by diseases that can spread via host-host in addition to long-distance spreading,
	in this work, we examine the effects of an additional local infection in the unidimensional CP with L\'evy flights.
	
	The remainder of this paper is organized as follows. In
	Sec. II we introduce the model and methods used in our analysis.
	In Sec. III we present our results. Section IV is
	devoted to discussion and conclusions.
	
	\section{Model and methods}
	
	To begin, we modify the standard contact process to include long distance infection. 
	An infected host, placed at a site $i$ on the lattice, is signed by a state variable $\sigma_i=1$. 
	It infects one of its healthy (signed by $\sigma_j=0$) nearest-neighbors with rate $\lambda_{h}$, or become healthy with rate $\mu=1$. In addition, the infected host can infect a healthy individual located at some distance, with rate $\lambda_{v}$. Here, $\lambda_{v}$ and $\lambda_{h}$ are the control parameters
	that govern the epidemic spreading. Each flight is described by using a L\'evy distribution \cite{levy1},
	which is characterized by an exponent $\alpha$. These events are schematically represented in Fig. 1. 
	Note that the evolution occurs in two independent steps, one concerning direct host-host infection,
	and the other, vector-mediated infection, related to the flights. 
	
	\begin{figure}[h]
		\begin{center}
			
			\begin{tikzpicture}[scale=0.5,rotate=0]
			\node at (-1,1) {\Large $t$:};
			\foreach \x in {0,...,9}
			\shadedraw [ball color= blue] (1.5*\x,1) circle (0.15cm);
			\shadedraw [ball color= red]  (3,1) circle (0.15cm);
			\shadedraw [ball color= red]  (4.5,1) circle (0.15cm);
			\shadedraw [ball color= blue]  (9,1) circle (0.15cm);
			\draw (2.85,1.25) edge[out=160,in=30,-latex, line width=1] (1.65,1.15);
			\node at (2.25,1.7) {$\lambda_h/2$};
			\draw (4.5,0.85) edge[out=270,in=90,-latex, line width=1] (4.5,-0.35);
			\node at (4.80,0.25) {$\mu$};
			\node at (-1.5,-0.5) {\Large $t+1$:};
			
			\foreach \x in {0,...,9}
			\shadedraw [ball color= blue] (1.5*\x,-0.5) circle (0.15cm);
			\shadedraw [ball color= red]  (3,-0.5) circle (0.15cm);
			\shadedraw [ball color= red]  (1.5,-0.5) circle (0.15cm);
			\shadedraw [ball color= blue]  (9,-0.5) circle (0.15cm);
			
			\node at (6.75,-1.5) {\Large +};
			\node at (-1,-3.5) {\Large $t$:};
			\foreach \x in {0,...,9}
			\shadedraw [ball color= blue] (1.5*\x,-3.5) circle (0.15cm);
			\shadedraw [ball color= red]  (3,-3.5) circle (0.15cm);
			\shadedraw [ball color= red]  (4.5,-3.5) circle (0.15cm);
			\shadedraw [ball color= blue]  (9,-3.5) circle (0.15cm);
			\draw (4.65,-3.35) edge[out=30,in=150,-latex, line width=1] (10.35,-3.35);

			\node at (-1.5,-5) {\Large $t+1$:};
			\foreach \x in {0,...,9}
			\shadedraw [ball color= blue] (1.5*\x,-5) circle (0.15cm);
			\shadedraw [ball color= red]  (3,-5) circle (0.15cm);
			\shadedraw [ball color= red]  (1.5,-5) circle (0.15cm);
			\shadedraw [ball color= blue]  (9,-5) circle (0.15cm);
			\shadedraw [ball color= red]  (10.5,-5) circle (0.15cm);
			\node at (7.5,-2.2) {$\lambda_v/2\ell$};
			
			\end{tikzpicture}
		\end{center}
		
		\caption {\label{fig:zica_rules} Transition rates for the model. Red circles represent infected and
			the blue ones represent the healthy individuals. {\em Top line:} local infection  (An infected individual transmites the disease to one of its first neighbors with rate $\lambda_h/2$, or become healthy with rate $\mu$.). {\em Bottom line:} long-distance infection (An infected individual transmites the disease to another individual with rate $\lambda_v/2\ell$, where $\ell$ is the distance between the individuals.)}
	\end{figure}
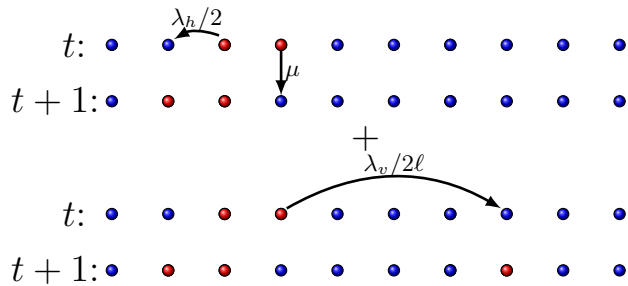
	
	This model can be interpreted as a prototypical model for diseases which can spread via
	vectors which can transmit the pathogen over long distances, and also by direct host-host
	contact. Since the CP is a spatial lattice model, it is in a core of models used in Ecology to investigate criticality in process determined by the local structure (configuration of trees) \cite{eco}.
	In nature, it is possible to occur long-range infections, as, for example in orchards where flying
	parasites contaminate the trees almost instantaneously
	in a widespread manner if the timescale of the
	flights of the parasites is much shorter than the mesoscopic
	timescale of the epidemic process itself \cite{janssen08}.  We should remark, however that our
	model is not suitable for modelling the spreading of human diseases such as the Zika virus, which is transmitted by mosquitoes, and 
	by parental or sexual contact \cite{zika1,zika2,zika3}. The length of intrinsic and extrinsic incubation periods of human vector-mediated diseases can be up to 2 weeks, and it is an essential factor that determines the transmission dynamics. In addittion, differently from trees, the human contact network is not a lattice, but instead a complex network, which also affects drastically the epidemic spreading \cite{networks1,networks2,networks3}. Both factors should be introduced in the model in a future work, if one intends modelling such diseases.
	
	In the simulation scheme, each step is divided in two events: (i) local, where an infected individual infects one of
	its neighbors (chosen at random) with probability $q=\lambda_{h}/(1+\lambda_{h})$ or it
	becomes healthy with probability  $\nu=1/(1+\lambda_{h})$; (ii) long-range, where an individual is infected with probability $q'=\lambda_{v}/(1+\lambda_{v})$,
	or nothing happens with probability  $v'=1/(1+\lambda_{v})$.
	To improve efficiency, we maintain a list of infected individuals, and time is updated
	as $\Delta t=1/N_{inf}$ at each iteration (here $N_{inf}$ is the number of
	infected individuals).
	
	The implementation of the vector flies consider a L\'evy-distributed random variables. The strategy to obtain such random flies is based on a nonlinear transformation of Gaussian random variables \cite{Janicki}. This strategy is based on the following recipe: {\it i.} First,  a random variable $V$ distributed homogeneously on ($-\pi/2, \pi/2$) is generated. {\it ii.} Second, an exponential variable $W$ with unity mean is generated. {\it iii.} Finally, we compute the random variable $\ell$~\cite{levy2}, following
	
	\begin{equation}
	\ell=\dfrac{\sin(\alpha V)}{[\cos(V)]^{\frac{1}{\alpha}}}\left\{\dfrac{\cos[(1-\alpha)V]}{W}\right\}^{\dfrac{1-\alpha}{\alpha}},
	\end{equation}
	where $\alpha$ is a control parameter of the length of the flies (this results in random flies with probability distribution function of length decaying as $\ell^{-(1+\alpha)}$) \cite{levy2}.
	
	We have done dynamical and stationary simulations. The dynamical simulations follow the system stochastic evolution from spreading (where 
	the system is initialized with a ``seed", i.e, only one site is infected at $t=0$), and initial decay studies (
	here the system evolves from a full infected lattice). 
	On the other hand, in the stationary simulations, we are interested in the (long-time) time independent behavior. 
	Stationary analysis nearby the critical point of systems with transitions into
	absorbing states are hard to be done due to strong finite size effects. This is consequence of that, in conventional simulations, small systems quickly become trapped in the absorbing state (in this case,
	we can say of a {\em quasistationary} state (since the only true stationary state is the absorbing one).
	In order to circumvent such difficulties, we employ a simulation method that yields
	quasistationary (QS) properties directly \cite{qssim1,qssim2}. The method
	is based in maintaining, and gradually updating, a list of $M$ configurations visited
	during the evolution; when a transition to the absorbing state is imminent, the system
	is instead placed in one of the saved configurations. Otherwise, the evolution is exactly
	that of a conventional simulation. This procedure results in an unbiased sampling
	of the quasistationary distribution of the process, and improve the statistics.
	
	\section{Results and discussion}
	
	In order to analyze the dynamical scaling of the model at criticality, we employed initial decay simulations in a unidimensional lattice of $L$ sites (with periodic boundary conditions). 
	The initial decay studies use an initial configuration with all sites occupied. The order parameter  $\rho(t)= \dfrac{1}{L} \sum \sigma_i$ is the density of active sites, i.e., the fraction of infected individuals.  At the critical point, one expects to observe a power-law behavior of the density 
	\begin{equation}
	\rho(t)=t^{-\delta},
	\end{equation} 
	until it saturates at its QS value. 
	The larger the system size, the
	longer the period of power-law decay, and the more precise the resulting estimate
	for the critical exponent $\delta$.
	In the initial decay studies, we have used systems with size up to $L=10^8$, and averages are taken over up to $10^6$ runs. 
	\begin{figure}[htb]
		\centering
		\includegraphics[width=.65\linewidth]{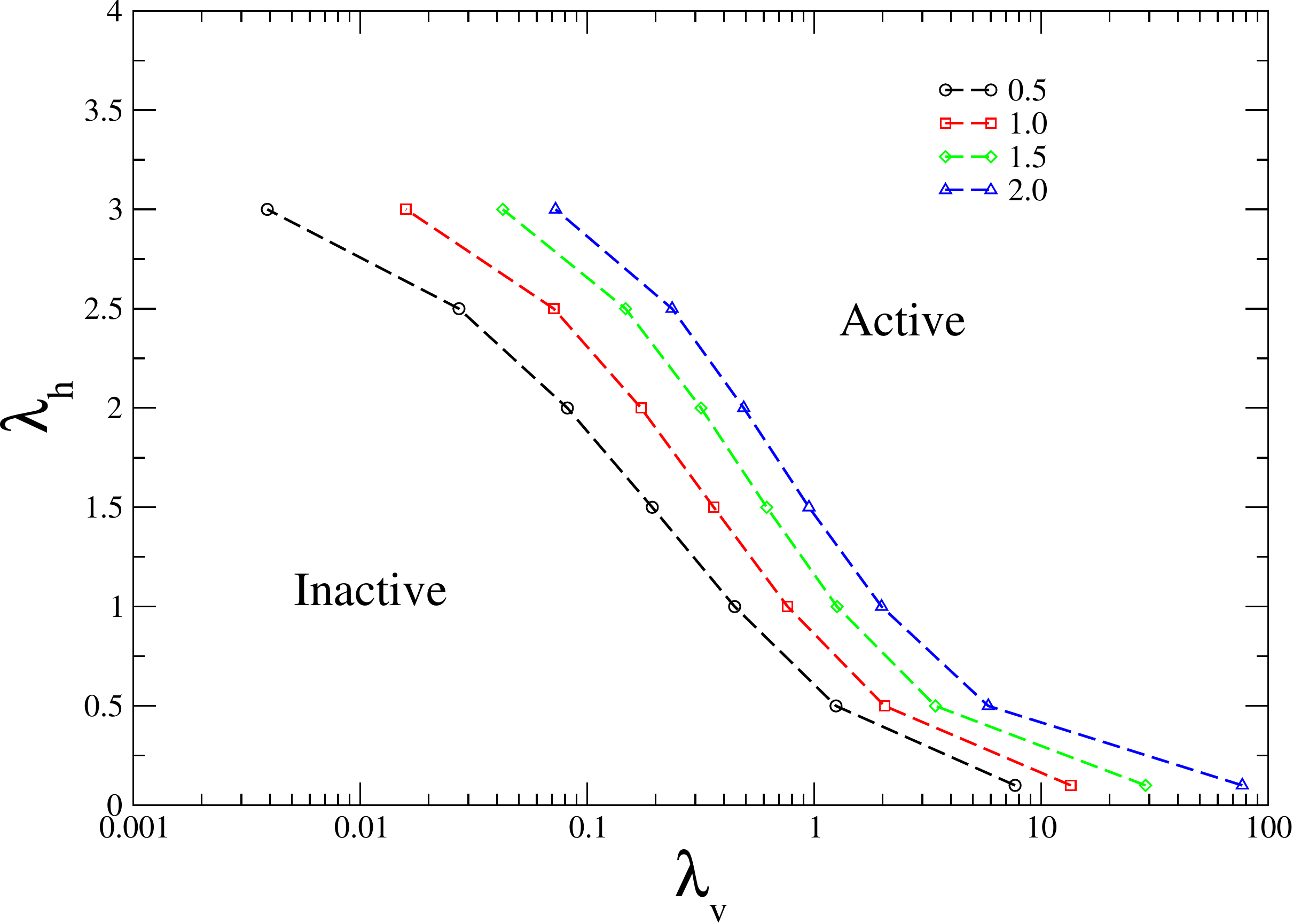}
		\caption{Phase diagram in the $\lambda_v \times \lambda_h$ plane, showing the inactive and active phases, for $\alpha=0.5,1.0,1.5$ and $2.0$ (from left to right).}
		\label{fig:phasediagram}
	\end{figure}
	
	From the spreading simulations, we observe there is a critical value of ($\lambda_h$,$\lambda_v$) above which the activity survives, while for values below such critical point, the system becomes trapped in the absorbing state. The phase diagram obtained in the $\lambda_h\times\lambda_v$ plane is plotted in Fig. 2. As expected, the absorbing phase is larger when $\alpha$ increases, since small values of $\alpha$ favor the epidemic spreading. 
	
	\begin{figure}[htb]
		\centering
		\includegraphics[width=0.6\linewidth]{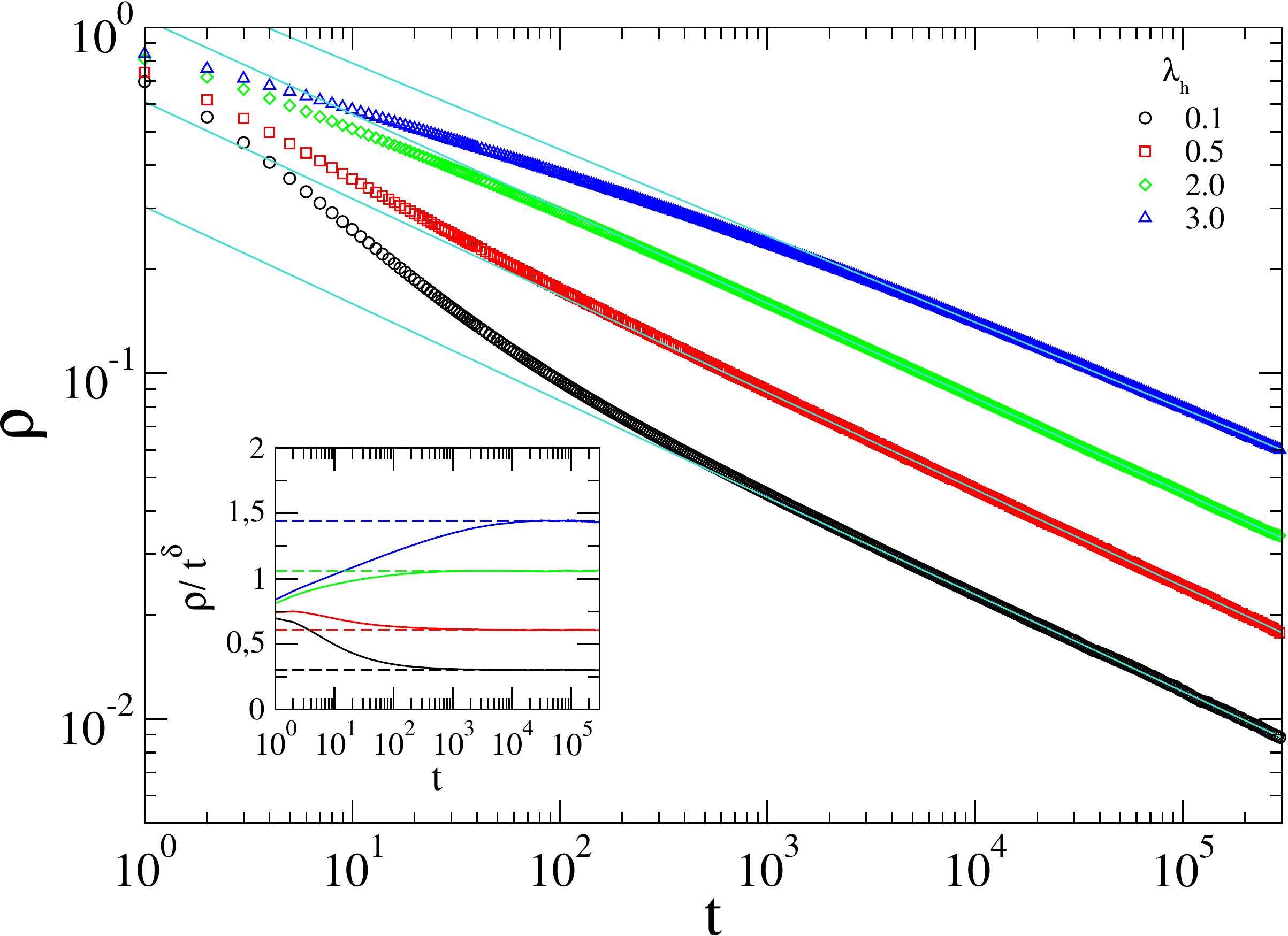}
		\caption{Decay of $\rho$  from an initial configuration with all sites infected  for distinct values of $\lambda_{h}$, for $\alpha=1.5$. 
			The straight lines are linear regressions from the data. The inset shows that $\rho / t^{\delta}$ converges to constant values in all cases, confirming the accuracy of the regressions.}
		\label{fig:rhottaxas}
	\end{figure}
	\begin{figure}[htb]
		\centering
		\includegraphics[width=0.6\linewidth]{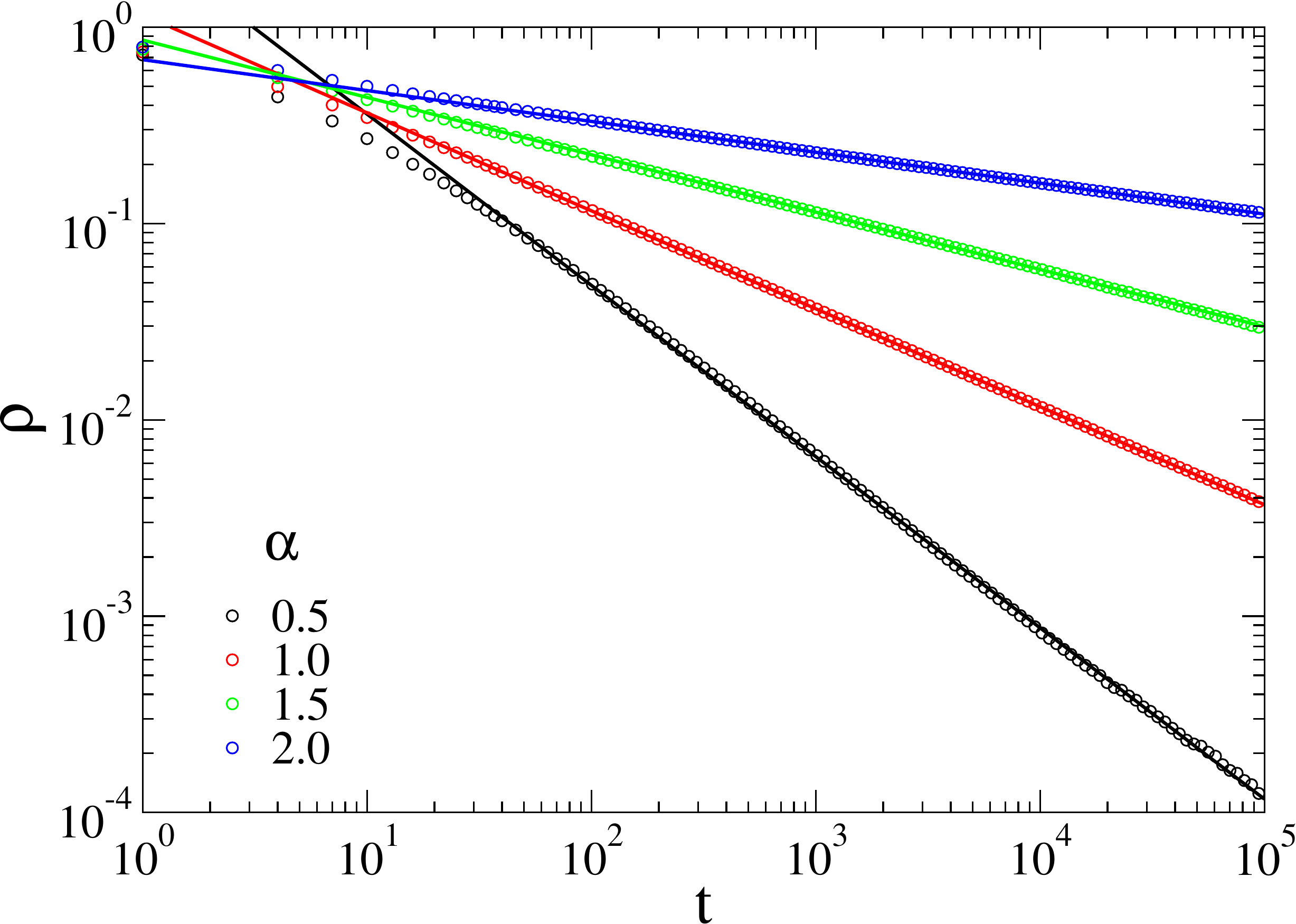}
		\caption{Decay of the density of infected sites $\rho$  from an initial configuration with all sites infected  for distinct values of $\alpha$, with $\lambda_{h}=1.0$. The straight lines are linear regressions from the data.}
		\label{fig:rhodecaialphas}
	\end{figure}
	
	In Fig. 3, we show the critical decay of the density of active sites, for distinct values of $\lambda_{h}$ and $\alpha=1.5$. In all cases, we observe a power law decay at criticality. Analysis from the data obtained yields the values of the critical exponent ($\delta$) reported in Table I. The inset in Fig. 3 shows that $\rho/ t^{\delta}$ converges to a constant value, therefore confirming the accuracy of the exponents obtained. We note the transient behavior is longer for larger values of $\lambda_{h}$. Also, the increase of  $\lambda_{h}$ alters only slightly the value of the exponent $\delta$.
	Figure 4 shows the evolution of the average density of infected sites for distinct values of $\alpha$. Increasing  $\alpha$, we observe a slower decay, as expected, since in such case, the flights are short-ranged. Increasing $\alpha$, the results become close to the original CP, with the critical exponent $\delta$ converging to the DP value,  $\delta=0.159$, as shown in Table I.    
	
	From the data in Table I, we observe that higher values of $\alpha$ recover the DP critical behavior, which presents $\delta=0.15947$ \cite{marro}. On the other hand, decreasing the value of $\alpha$, the exponent $\delta$ approaches the mean-field value $\delta=1$. The critical exponent $\delta$ varies continuously between these limits. We observe that the effects of $\lambda_h$ are more pronounced
	for lower values of $\alpha$, since the exponent $\delta$ varies from $0.95$ to $0.82$ when $\alpha=0.5$. We note, however, that $\alpha=0.5$ is  a crossover point, where the dynamical behavior is often plagued by huge corrections to scaling. Otherwise, for higher values of $\alpha$  the exponent does not alter significantly.
	Finally, in comparison with the results for the CP with only L\'evy-type
	infection (i.e, with $\lambda_h=0.0$) \cite{hinri99}, we observe the nonuniversal critical exponents are closer the DP values (for the same value of $\alpha$) when the host-host infection is introduced. For example, in \cite{hinri99} it was found $\delta=0.21$ for
	$\alpha=2.0$ and $\delta=0.94$ for $\alpha=0.5$, in contrast to $\delta=0.16$ and $\delta=0.82$, respectively, in the present work with $\lambda_h=3.0$. 
	
	\begin{table}[htb]
		\begin{center}
			\begin{tabular}{c|cc|cc|cc|cc} 
				
				\hline \hline 
				& \multicolumn{8}{c}{$\alpha$}  \\
				\hline
				& \multicolumn{2}{c|}{$0.5$}& \multicolumn{2}{c|}{$1.0$} & \multicolumn{2}{c|}{$1.5$}   & \multicolumn{2}{c}{$2.0$}  \\ 
				\hline \hline 
				\rule[-1ex]{0pt}{2.5ex} $\lambda_{h}$ & { $\lambda_{v}$} & $\delta$ & $\lambda_{v}$ & $\delta$& $\lambda_{v}$ & $\delta$& $\lambda_{v}$ & $\delta$ \\
				\hline
				
				\rule[-1ex]{0pt}{2.5ex} 0.1 & 7.68415(5) & 0.94(1) &13.5159(5)   & 0.50(1) & 28.8596(5) & 0.29(2) & 77.98(1)   & 0.16(1) \\ 
				
				\rule[-1ex]{0pt}{2.5ex} 0.5 & 1.24847(5) & 0.91(1) & 2.0433(1)   & 0.51(2) & 3.4256(1)  & 0.28(1) & 5.8545(5)   & 0.16(1) \\ 
				
				\rule[-1ex]{0pt}{2.5ex} 1.0 & 0.446503(5)& 0.89(1) & 0.764393(5) & 0.49(1) & 1.2614(1)  & 0.29(1)& 1.9832 (5) & 0.16(1) \\ 
				
				\rule[-1ex]{0pt}{2.5ex} 1.5 & 0.19355(1) & 0.89(1) & 0.36126(1)  & 0.49(1) & 0.6185(1)  & 0.26(1) & 0.9578(1)  & 0.16(1) \\ 
				
				\rule[-1ex]{0pt}{2.5ex} 2.0 & 0.081827(1)& 0.89(2) & 0.1728(1)   & 0.49(2) & 0.3177(1)  & 0.26(1) & 0.4972(1)   & 0.16(1) \\ 
				
				\rule[-1ex]{0pt}{2.5ex} 2.5 & 0.027251(1)& 0.85(2) & 0.07132(1)  & 0.49(2) &  0.1477(1) & 0.26(1) & 0.2378(1)   & 0.16(1) \\ 
				
				\rule[-1ex]{0pt}{2.5ex} 3.0 & 0.003903(1)& 0.82(2) & 0.015922(5) & 0.48(2) &  0.0425(1) & 0.23(2) & 0.07258(1) & 0.16(1) \\ 
				\hline \hline 
				
			\end{tabular} 
			
		\end{center}
		\caption {Critical values of the control parameter, $\lambda_{v}^*$, and critical exponent $\delta$ as a function of $\lambda_h$ for $\alpha=0.5, 1.0, 1.5$ and $2.0$.}
	\end{table}
	
	Now, we turn to the static, long-time behavior, represented by the quasi-stationary state. In the QS simulations, we have used system sizes  ranging from $L=10^2$ to $L=10^{5}$. Each simulation ran until $t=10^8$, and averages were taken over $10^3$ runs. We used a list of size $M=2000$. At the critical point, the finite-size theory \cite{marro} implies the quasistationary order
	parameter $\rho_{qs}$ decays with the system size as a power law 
	\begin{equation}
	\rho_{qs}\sim L^{-\beta/\nu_\perp}.	
	\end{equation}
	
	In Fig. 5, we show the QS density of infected sites  ($\rho_{qs}$) as a function of the system size $L$, for $\alpha=1.5$. We observe that at criticality, $\rho_{qs}$ scales as a power law with exponent ratio $\beta/\nu_\perp=0.45$, $0.42$, $0.42$, $0.43$, $0.42$, $0.41$ and $0.44$ for $\lambda_h=0.1,~0.5,~1.0,~1.5,~2.0,~2.5$ and $3.0$, respectively. Hence, we can conclude that the exponent $\beta/\nu_\perp$ is independent of $\lambda_h$.
	\begin{figure}[htb]
		\centering
		\includegraphics[width=0.6\linewidth]{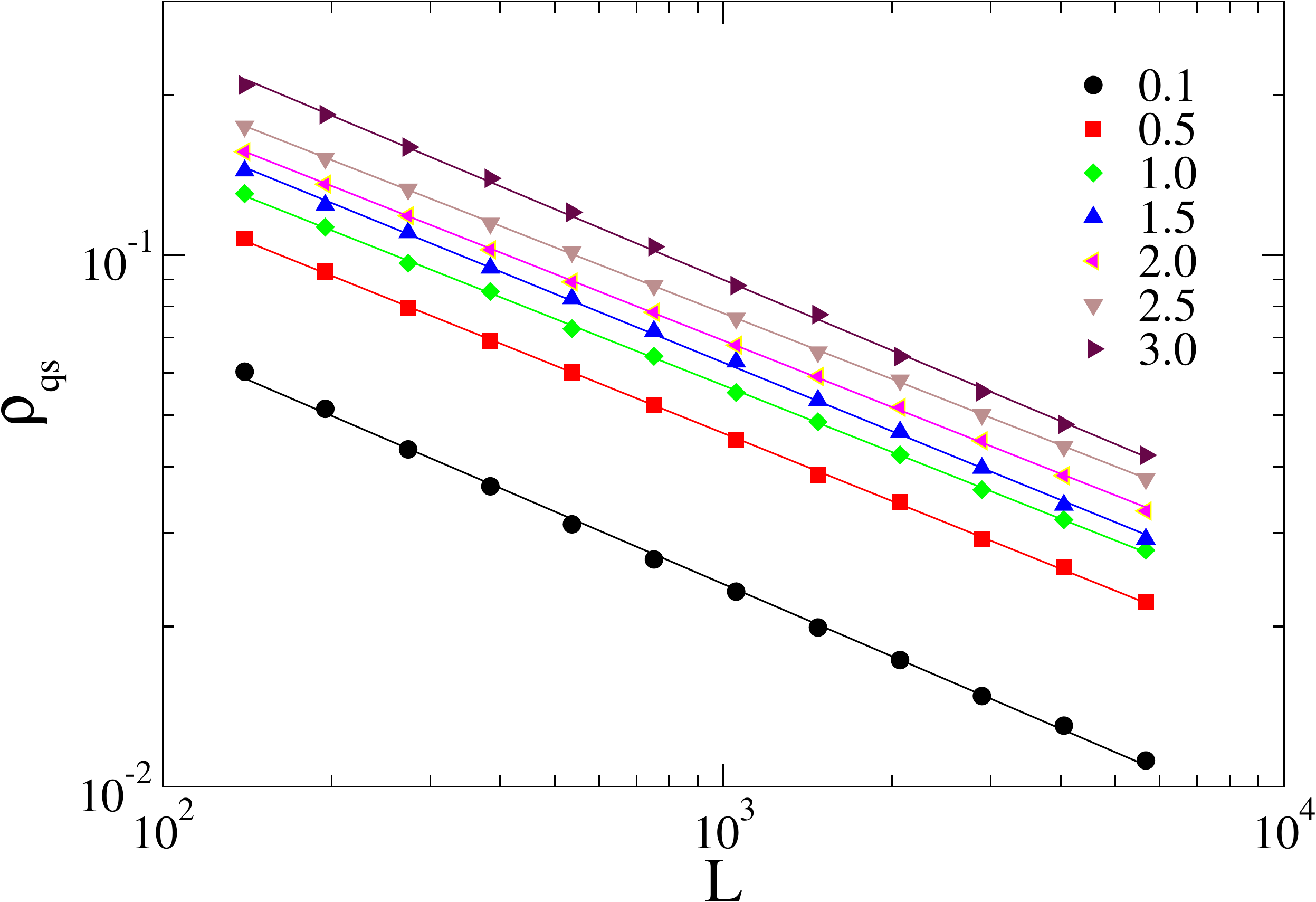}
		\caption{Finite size scaling of the QS density $\rho_{qs}$, for $\alpha=1.0$ and $\lambda_h=0.1,~0.5,~1.0,~1.5,~2.0,~2.5$ and $3.0$. }
		\label{fig:rhoqslh1ls}
	\end{figure}
	
	In Fig. 6, we evaluate the behavior of the lifetime $\tau$ of the QS state at the criticality, which is expected to scale as
	\begin{equation}
	\tau \sim L^{z},
	\end{equation}
	with $z=\nu_\parallel/\nu_\perp$ (here, we evaluate the lifetime as the time between two consecutive visits to the absorbing state). Analysis from the data
	yields $z=0.84$, $0.83$, $0.83$, $0.80$, $0.82$, $0.84$ and $0.83$ for $\lambda_h=0.1,~0.5,~1.0,~1.5,~2.0,~2.5$ and $3.0$, respectively. Therefore, this exponent also is not 
	affected by the host-host interaction. Results of the QS simulations for distinct values of $\alpha$ are reported in Table II. We observe that in all cases the critical exponents obtained from the QS simulations are not significantly affected by the host-host infection.  
	
	In resume, we conclude that the introduction of a local host-host interaction in epidemics with L\'evy-type long-range interactions does not affect the long-time critical behavior. On the other hand, we observe an anomalous spreading observed for the dynamical critical exponent for $\alpha=0.5$.

	\begin{figure}[htb]
		\centering
		\includegraphics[width=0.6\linewidth]{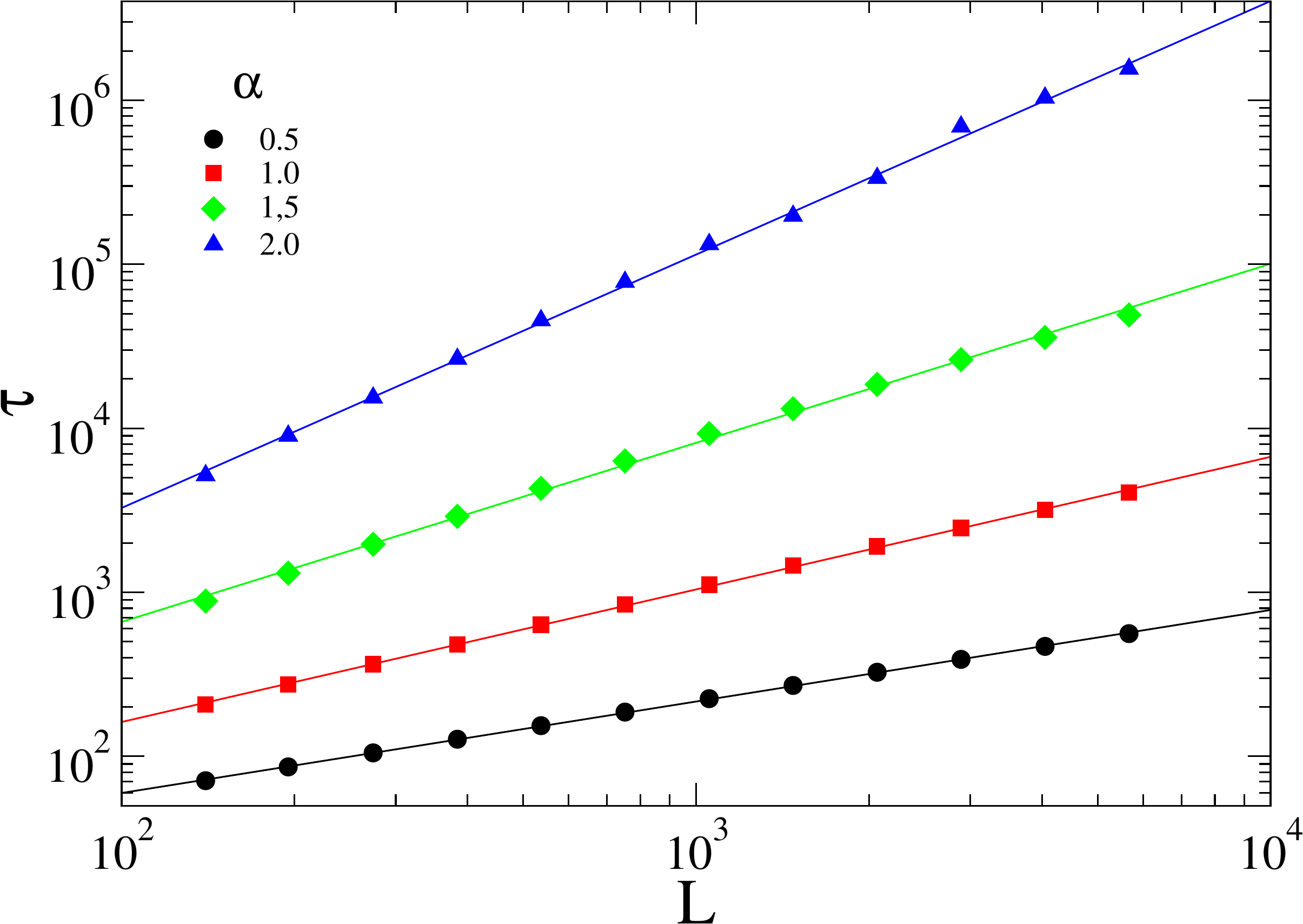}
		\caption{Finite size scaling of the QS lifetime $\tau_{qs}$, for $\alpha=1.5$ and $\lambda_h=0.5,1.0,2.0$ and $3.0$. }
		\label{fig:tau_L}
	\end{figure}
	
	\begin{table*}
		\begin{center}
			\begin{tabular}{c|cc|cc|cc|cc} 
				
				\hline \hline
				& \multicolumn{8}{c}{$\alpha$}  \\
				\cline{2-9} 
				& \multicolumn{2}{c}{$0.5$}& \multicolumn{2}{c}{$1.0$} & \multicolumn{2}{c}{$1.5$}   & \multicolumn{2}{c}{$2.0$}  \\ 
				\hline \hline 
				\rule[-1ex]{0pt}{2.5ex} $\lambda_{h}$ & $\beta/\nu_\perp$ & $z$ & $\beta/\nu_\perp$ & $z$& $\beta/\nu_\perp$ & $z$ & $\beta/\nu_\perp$ & $z$ \\
				\hline
				
				\rule[-1ex]{0pt}{2.5ex} 0.1 & 0.55(2) & 0.56(1) & 0.45(1) & 0.84(2) & 0.36(2) & 1.20(3) & 0.26(2)  & 1.60(3) \\ 
				
				\rule[-1ex]{0pt}{2.5ex} 0.5 & 0.52(2) & 0.56(1) & 0.42(2) & 0.82(2) & 0.35(2) & 1.18(4) & 0.26(2) & 1.53(2)\\ 
				
				\rule[-1ex]{0pt}{2.5ex} 1.0 & 0.51(1) & 0.56(1) & 0.42(2) & 0.82(1) & 0.34(1) & 1.16(4) & 0.25(2) & 1.56(4)  \\ 
				
				\rule[-1ex]{0pt}{2.5ex} 1.5 & 0.50(1) & 0.56(1) & 0.43(1) & 0.80(1) & 0.35(2) & 1.09(3) & 0.26(3) & 1.60(3) \\ 
				
				\rule[-1ex]{0pt}{2.5ex} 2.0 & 0.49(1) & 0.58(1) & 0.42(2) & 0.82(1) & 0.34(3) & 1.14(3) & 0.23(4) & 1.67(3) \\ 
				
				\rule[-1ex]{0pt}{2.5ex} 2.5 & 0.50(2) & 0.58(2) & 0.42(2) & 0.83(2) & 0.33(2) & 1.17(3) & 0.26(3) & 1.53(3) \\ 
				
				\rule[-1ex]{0pt}{2.5ex} 3.0 & 0.50(2) & 0.60(2) & 0.43(2) & 0.83(2) & 0.34(6) & 1.24(3) & 0.27(3) & 1.52(3) \\ 
				\hline \hline 
				
			\end{tabular} 
		\end{center}
		\caption {Critical values of the control parameter, $\lambda_{v}^*$, and critical exponent $\delta$ as a function of $\lambda_h$ for $\alpha=0.5, 1.0, 1.5$ and $2.0$.}
	\end{table*} 
	
	\section{Conclusions}
	
	We have proposed a variation of the contact process that includes both local (host-host) and long-distance (vector-mediated) interactions.  
	We observe the host-host interactions does not alter the static and dynamical critical exponents  significantly in comparison to a process with only L\'evy-type interactions. Our
	results are in agreement with early field-theoretic results \cite{janssen08}, which revealed that the relative strength of short-and long-range interactions do not affect the critical behavior.
	Our results also show an anomalous spreading, with the dynamical exponent varying continuously with the host-host infection rate, when $\alpha=0.5$. However, we should remark 
	that at this crossover point, the dynamical behavior is affected by huge corrections to scaling, and we cannot discard that this difference could vanish for sufficient long times (note that 
	the static simulations show the same set of exponents for all values of $\alpha$).  
	Finally, in the limit in which the host-host infection vanishes, our model reduces to the anomalous contact process proposed in \cite{hinri99}, and our results in this limit confirm the critical behavior
	obtained previously in the literature. 
	
	A promising extension of the present work includes the study of the model under the influence of temporal disorder \cite{tempd,tempd2}, that could, for example, affect only the
	vector-mediated infection, reflecting seasonal variations \cite{dengue}. Another critical issue is the effect of local interaction in the robustness of the discontinuous phase transition exhibited by some classes of long-range models \cite{ginelli05,ginelli06,firstorder}.

	\section*{Conflict of Interest Statement}
	
	The authors declare that the research was conducted in the absence of any commercial or financial relationships that could be construed as a potential conflict of interest.
	
	\section*{Author Contributions}
	
	SGA and MMO conceived the study, PRS and SGA performed the simulations, PRS, SGA and MMO analyzed the data, PRS, SGA and MMO wrote the paper. All authors reviewed the manuscript.
	
	\section*{Aknowledgments}
	
	This work was partially supported by Brazilian agencies CNPq and FAPEMIG.
	
	\vspace{3cm}
	
	\bibliographystyle{frontiersinHLTH&FPHY}

\begin{thebibliography}{100}
		
		\bibitem{harris-CP}
		T.~E. Harris, Contact interactions on a lattice,
		Ann. Probab., {\bf 2}, 969 (1974).
		
		\bibitem{durrett}
		R. Durrett, Stochastic spatial models,
		SIAM Rev. {\bf 41}, 677 (1994).
		
		\bibitem{marro}
		J. Marro and R. Dickman,
		{\it Nonequilibrium Phase Transitions in Lattice Models}
		(Cambridge University Press, Cambridge, 1999).
		
		\bibitem{odor07}
		G. \'Odor,
		{\it Universality In Nonequilibrium Lattice Systems: Theoretical Foundations}
		(World Scientific,Singapore, 2007)
		
		\bibitem{henkel}
		M. Henkel, H. Hinrichsen and S. Lubeck,
		{\it Non-Equilibrium Phase Transitions Volume I: Absorbing Phase Transitions}
		(Springer-Verlag, The Netherlands, 2008).
		
		\bibitem{hinrichsen}
		H. Hinrichsen, Non-equilibrium critical phenomena and
		phase transitions into absorbing states, 
		Adv. Phys. {\bf 49}, 815 (2000).
		
		\bibitem{odor04}
		G. \'Odor, Universality classes in nonequilibrium lattice
		systems,
		Rev. Mod. Phys {\bf 76},  663 (2004).
		
		\bibitem{zgb} R. M. Ziff, E. Gulari, and Y. Barshad, Kinetic phase transitions
		in an irreversible surface-reaction model, Phys. Rev. Lett. 56, 2553 (1986).
		
		\bibitem{tang} L. H. Tang and H. Leschhorn, Pinning by directed percolation, Phys. Rev. A 45, R8309(1992).
		
		\bibitem{scp}
		M.~M. de Oliveira, R.~V. Santos and R. Dickman, Symbiotic two-species contact process,
		Phys. Rev. E {\bf 86}, 011121 (2012).
		
		\bibitem{take07}
		K.~A. Takeuchi, M. Kuroda, H. Chat\'e, and M. Sano, Directed
		percolation criticality in turbulent liquid crystals,
		Phys. Rev. Lett. {\bf 99}, 234503 (2007).
		
		\bibitem{pine}
		L. Cort\'e, P. M. Chaikin, J. P. Gollub, and D. J. Pine, Random organization in periodically driven systems.
		Nature Physics {\bf 4}, 420 (2008).
		
		\bibitem{okuma} S. Okuma, Y. Tsugawa, and A. Motohashi, Transition
		from reversible to irreversible 
		flow: Absorbing and depinning
		transitions in a sheared-vortex system,
		Phys. Rev. B{\bf 83}, 012503 (2011).
		
		\bibitem{janssen}
		H.~~ K. Janssen, On the nonequilibrium phase transition in reaction-diffusion systems with an absorbing stationary
		state,
		Z. Phys. B {\bf 42}, 151 (1981).
		
		\bibitem{grassberger}
		P. Grassberger, On phase transitions in schloogl\'s second
		model,
		Z. Phys. B {\bf 47}, 365 (1982).
		
		\bibitem{dengue}  T. Botari, S.G. Alves, E.D. Leonel, Explaining
		the high number of infected people by dengue in Rio
		de Janeiro in 2008 using a susceptible-infective-recovered
		model,
		Phys. Rev. E 83, 037101 (2011).
		
		\bibitem{grass86} P. Grassberger, in Fractals in physics, edited by L.
		Pietronero, E. Tosatti (Elsevier, 1986).
		
		\bibitem{mollison} D. Mollison, Spatial contact models for ecological and
		epidemic spread, J. R. Stat. Soc. Ser. B Methodol. {\bf 39}, 283 (1977).
		
		\bibitem{hinri99} H. Hinrichsen and M. Howard, A model for anomalous
		directed percolation, Eur. Phys. J. B {\bf 7}, 635 (1999).
		
		\bibitem{hinri07} H. Hinrichsen, Non-equilibrium phase transitions with long-range interactions, J. Stat. Mech.: Theory Exp. 2007, P07006.
		
		\bibitem{janssen99} H. K. Janssen, K. Oerding, F. van Wijland, and H. J. Hilhorst, L\'evy-flight spreading of epidemic processes
		leading to percolating clusters,
		Eur. Phys. J. B {\bf 7}, 137 (1999).
		
		\bibitem{ginelli05} F. Ginelli, H. Hinrichsen, R. Livi, D. Mukamel, and A. Politi, Directed percolation with long-range interactions:
		Modeling nonequilibrium wetting,
		Phys. Rev. E {\bf 71}, 026121 (2005).
		
		\bibitem{ginelli06} F. Ginelli, H. Hinrichsen, R. Livi, D. Mukamel, and A. Torcini, Contact processes with long range interactions,
		J. Stat. Mech.: Theory Exp. {\bf 2006}, P08008 (2006).        
		
		
		\bibitem{fiore07} C. E. Fiore and M. J. de Oliveira, Contact process with
		long-range interactions: A study in the ensemble of constant
		particle number, Phys. Rev. E {\bf 76}, 041103
		(2007).
		
		\bibitem{fiore13} C. E. Fiore and M. J. de Oliveira, Robustness of first order
		phase transitions in one-dimensional long-range contact processes,
		Phys. Rev. E {\bf 87}, 042101 (2013).
		
		\bibitem{igloi} R. Juh\'asz, I. A. Kov\'acs, and F. Igl\'oi, Long-range epidemic
		spreading in a random environment,
		Phys. Rev. E {\bf 91}, 032815 (2015).
		
		\bibitem{pedro} T. B. Pedro, W. Figueiredo, and A. L. Ferreira, Mean-field theory for the long-range contact process with diffusion
		Phys. Rev. E {\bf 92}, 032131 (2015).
		
		\bibitem{levy1}  E.W. Montroll, B.J. West, Fluctuation Phenomena, edited
		by E.W. Montroll, J.L. Lebowitz (North-Holland, Amsterdam,
		1979).
		
		\bibitem{eco} M. Pascual, F. Guichard, Criticality and disturbance in spatial
		ecological systems, Trends Ecol. Evol. {\bf 20}, 88
		(2005).
		
		\bibitem{janssen08} H.-K. Janssen and O. Stenull, Field theory of directed percolation with long-range spreading,
		Phys. Rev. E {\bf 78}, 061117 (2008). 
		
		\bibitem{zika1} E. D'Ortenzio {\it et al.}, Evidence of sexual transmission of zika virus,
		N. Engl. J. Med. {\bf 374} , 2195 (2016).
		
		\bibitem{zika2} F. C. Coelho {\it et al.}, Higher
		incidence of zika in adult women than adult men in rio
		de janeiro suggests a significant contribution of sexual
		transmission from men to women, Int. J. of Infectious Diseases {\bf 51}, 128 (2016).
		
		\bibitem{zika3} S. Towers {\it et al.}, Estimate of the
		reproduction number of the 2015 zika virus outbreak in
		Barranquilla, Colombia, and estimation of the relative role
		of sexual transmission, Epidemics {\bf 17}, 50 (2016).
		
		\bibitem{Janicki} A. Janicki and A. Weron, Computer simulation of attractors in stochastic models with $\alpha$-stable noise, 
		Mathematics and Computers in Simulation {\bf 39}, 9 (1995).
		
		\bibitem{levy2} J. Klafter and M. Sokolov. \textit{First Steps in Random Walks}. Oxford University Press, New York, 2011.
		
		\bibitem{networks1} C. Moore and M. E. J. Newman, Epidemics and percolation in small-world networks,
		Phys. Rev. E {\bf 61}, 5678 (2000).
		
		\bibitem{networks2} R. Parshani, S. Carmi and S. Havlin, Epidemic Threshold for the Susceptible-Infectious-Susceptible Model on Random Networks,
		Phys. Rev. Lett. {\bf 104}, 258701 (2010).
		
		\bibitem{networks3} R. Pastor-Satorras, C. Castellano, P. Van Mieghem, and A. Vespignani, Epidemic processes in complex networks,
		Rev. Mod. Phys. {\bf 87}, 925 (2015).
		
		\bibitem{qssim1}
		M.~M. de Oliveira and R. Dickman, How to simulate the
		quasistationary state, Phys. Rev. E {\bf 71}, 016129 (2005).
		
		\bibitem{qssim2}  R. Dickman and M.~M. de Oliveira, Quasi-stationary simulation: The subcritical contact process,
		Braz. J. of Physics 36, 685 (2006).
		
		\bibitem{tempd} M. M. de Oliveira and C. E. Fiore, Temporal disorder
		does not forbid discontinuous absorbing phase transitions
		in low-dimensional systems, Phys. Rev. E {\bf 94}, 052138 (2016).
		
		\bibitem{tempd2} C. M. D. Solano, M. M. de Oliveira and C. E. Fiore, Comparing the influence of distinct kinds of temporal
		disorder in a low-dimensional absorbing transition model,
		Phys. Rev. E {\bf 94}, 042123 (2016).
		
		\bibitem{firstorder}M.M. de Oliveira, M.G.E da Luz, C.E. Fiore, Generic finite size scaling for discontinuous nonequilibrium
		phase transitions into absorbing states, Phys. Rev. E {\bf 92}, 062126 (2015).
		
	\end{thebibliography}

	\newpage

\end{document}